 \scrollmode
 \documentstyle[11pt]{article}
\begin{document}
 \title{{\bf Intermittency in the $q$-state Potts model\vspace{1cm}}}
 \author{Yves LEROYER\vspace{1cm}\\
{\em Laboratoire de Physique Th\'eorique } \\
Unit\'e Associ\'ee au CNRS, U.A. 764 \\
{\em Universit\'e de Bordeaux I} \\
{\em Rue du Solarium }\\
{\em F-33175 Gradignan Cedex}}
\date{ }
\begin{titlepage}
  \maketitle
\thispagestyle{empty}
\begin{center}
{\large {\bf Abstract}}\vspace{0.5cm}
\end{center}
We define a block observable for the $q$-state Potts model which
 exhibits an intermittent behaviour  at the critical point. We express the
intermittency indices of the normalised moments in terms of the magnetic
critical exponent $\beta /\nu$ of the
model. We confirm this relation by a numerical similation of the $q=2$ (Ising)
and $q=3$ two-dimensional Potts model.  \vspace{2.0cm}\\

\noindent LPTB 93-2\\
Mars 1993
e-mail : LEROYER@FRCPN11.IN2P3.FR
\end{titlepage}
\newpage

\section{Introduction}
Recently, the concept of intermittency has been intoduced in the field of
equilibrium critical phenomena as a tool for studying local fluctuations of a
system undergoing a second order phase transition~\cite{Pesch,Wosiek}. For
example, in the Ising model, as the
the magnetisation fluctuates without limit at the critical point,
one expects that the normalised moments of this observable, measured on
sub-blocks of
the system, scale with the size of the block, with exponents that may depend on
the order of the moment, the so-called intermittency indices.
On the basis of real space renormalisation group arguments,  Satz~\cite{Satz}
established that, for the Ising
model, the intermittency indices are directly connected to the magnetic
critical exponent. However, the normalised moments of the magnetisation used in
Satz's argument
are not suitable for a numerical simulation, as the net magnetisation, which
enters in the denominator of  the moments,  is almost zero
on a (large) finite lattice.
In subsequent numerical simulations, Bambah et. al.\cite{Bambah}
and Gupta et. al.\cite{Gupta},	using a new observable which respects the
$Z_2$ symmetry and remains finite on finite lattices, present clear evidence
for an intermittent behaviour  in the two-dimensional Ising model. However,
they do not observe the
expected relation between the intermittency exponents and the critical ones.
Burda et. al.~\cite{Zalewski}  suggest that this is a consequence of their
choice
for the fluctuating observables, among which the moments have only a
subdominant scaling behaviour in the critical regime.

Therefore, the problem of finding convenient block observables which have
well-defined moments on finite lattices and
which exhibit the expected scaling law,  remains open. In
this paper we address this question in the more general framework of the
$q$-state Potts model.

 In section 2, we define the fluctuating observable
for which we derive the intermittency
indices in terms of the magnetic exponent by generalising the Satz's
argument~\cite{Satz}. In section 3 we give the details of the numerical
simulation and discuss the results. Conclusions are drawn in the last section.

\section{Definition of the observables}
The most convenient definition of the moments for studying critical spin
systems has been thoroughly discussed
in  ref~\cite{Pesch,Gupta}. Following the authors of ref.~\cite{Gupta} we use
the so-called {\em standard block} moments of order $p$ defined as
\begin{equation}
F_p(L;\ell )=\frac{1}{M}\sum_\alpha \frac{<k_\alpha^p>}{<k_\alpha >^p}
\end{equation}
where $L$ is the lattice size, $\ell$ the cell size,  $M=(L/\ell )^d$
the number of cells and $k_\alpha$ the block observable
defined on the $\alpha^{th}$ cell. The symbols $<>$ stands for the
thermodynamical average which will be taken
{\em at the bulk critical temperature} $T_c$ defined by $K_c=J/k_bT_c=\ln
[1+\sqrt{q}]$ for the $q$-state Potts model. We recall that the intermittency
indices $\lambda_p$ are defined according to the behaviour of the moments with
respect to the block size $\ell$~:
$$F_p(L;\ell )\sim \left(\frac{\ell}{L}\right)^{-\lambda_p}\qquad \mbox{for}\;
\frac{\ell}{L}\rightarrow 0$$

Let us now define the block observable $k_\alpha$.

We consider a system of Potts spins $s_i=0,1,\ldots,q-1$, interacting
with their nearest neighbours via the hamiltonian
$$-\beta H =K\sum_{<i,j>}\delta_{s_is_j}$$
on a two-dimensional square lattice of linear size $L$.
For a particular configuration, we denote by $Q$ the spin value which occurs
the most frequently and to each site $i$ we assign the variable
$n_i=\delta_{s_iQ}$. We now define
$$k_i= \frac{qn_i-1}{q-1}$$
which is such that
$$\frac{1}{L^d}<\sum_i k_i>=<k_i>=<m_q>$$
is the order parameter for the Potts model~\cite{Wu}. For the Ising model
($q=2$) this definition reduces to
$$k_i=\mbox{sgn}(m_I)\sigma_i$$
where $m_I$ is the Ising magnetisation of the configuration of Ising spins
$\{\sigma_i=\pm 1\}$.\\
We then define the {\em block} observable to be
\begin{equation}
k_\alpha =\frac{1}{\ell^d}\sum_{i\in \alpha} k_i =  \frac{qn_Q^{(\alpha
)}-1}{q-1}
\end{equation}
where  $n_Q^{(\alpha )}$ is the fraction
of spins in the cell $\alpha$ having the value $Q$~\cite{Hajdu}.

Unlike the original definition of ref~\cite{Satz},
 on a {\em finite lattice}, $<k_\alpha >$ is non zero at the
critical point, ($<k_\alpha >=<|m_I|>$ for the Ising model) allowing the
moments of eq.~(2.1) to be measured in a numerical simulation.
Moreover, one can extend to this observable the renormalisation argument which
predicts the intermittent behaviour of the moments.\\
 Let us associate to the blocking procedure the following renormalisation
prescription~: we define a block spin $s_\alpha$ according to the majority rule
in the $\alpha^{th}$ cell, associate the corresponding renormalised site
variable\footnote{For a given
configuration, the value of $Q$ is preserved by the blocking procedure}
 $\tilde{n}^{(\alpha )}_Q=\delta_{s_\alpha Q}$ and then the renormalised
quantity $\tilde{k}_\alpha$
 $$\tilde{k}_\alpha = \frac{q\tilde{n}_Q^{(\alpha )}-1}{q-1}$$
Since $k_\alpha$ is related to the magnetic scaling operator, we expect the
following renormalisation relation
$$k_\alpha = Z(\ell )\tilde{k}_\alpha$$
where $Z(\ell )$ is the renormalisation factor depending only on the block
size $\ell$. Therefore
\begin{eqnarray}
<k_\alpha^p> & = & \left[Z(\ell )\right]^p<\left(\frac{q\tilde{n}_Q^{(\alpha
)}-1}{q-1}\right)^p>\\
 & =  &  \left[Z(\ell )\right]^p \left[ <\tilde{n}_Q^{(\alpha )}>\left(
1+\frac{(-1)^{(p+1)}} {(q-1)^p} \right)+\frac{(-1)^p}{(q-1)^p}\right]
\end{eqnarray}
where we have used $(\tilde{n}_Q^{(\alpha )})^p=\tilde{n}_Q^{(\alpha )}$.
Reintroducing $\tilde{k}_\alpha$ in the left-hand side of eq.(2.4),
 we get the relation \begin{equation}
<k_\alpha^p> = \left[Z(\ell )\right]^p (A_p <\tilde{k}_\alpha >+B_p)
\end{equation}
where the coefficients $A_p$ and $B_p$ are given exactly by
\begin{eqnarray}
A_p & =  & \frac{q-1 }{q} \left( 1+\frac{(-1)^{(p+1)}} {(q-1)^p} \right)\\
B_p & =  & \frac{1 }{q} \left( 1+\frac{(-1)^{(p+1)}} {(q-1)^p} \right)
+\frac{(-1)^p}{(q-1)^p}
\end{eqnarray}

According to the standard renormalisation analysis, we expect
$Z(\ell )\sim \ell^{-\frac{1}{2}(d-2+\eta )} = \ell^{-\beta /\nu}$. Moreover,
$<k_\alpha >$ and $<\tilde{k}_\alpha>$ are the order parameter of
the system of size $L$ and of the
rescaled system of size $L/\ell$, respectively. At the bulk critical
temperature,
these quantities behave according to the finite size scaling law
\begin{eqnarray}
<k_\alpha > & \sim & L^{-\beta/\nu}\\
<\tilde{k}_\alpha > & \sim & (L/\ell)^{-\beta/\nu}
\end{eqnarray}

Thus, for $\ell\gg 1$ (in lattice units)
 in order to sum up enough degrees of freedom for the renormalisation
arguments to be valid and $\ell \ll L$ to avoid finite size effects, we expect
the moments $F_p$ to behave like
\begin{equation}
F_p(L;\ell )   \sim   \left(\frac{\ell}{L}\right)^{-p\frac{\beta}{\nu}}
\left[A'_p \left(\frac{\ell}{L}\right)^{\frac{\beta}{\nu}} +B'_p\right]
\end{equation}
where the coefficients $A'_p$ and $B'_p$ are proportional to the coefficients
$A_p$ and $B_p$ defined above. \\

For the Ising model $(q=2$), eqs (2.6) and (2.7) give $A_p\equiv 0$ for $p$
even and $B_p\equiv
0$ for $p$ odd, leading to the scaling law, derived previously by
Satz~\cite{Satz} \begin{eqnarray}
F_p(L;\ell ) & \sim & \left(\frac{\ell}{L}\right)^{-p\frac{\beta}{\nu}}
\qquad \mbox{for $p$ even}\\
  & \sim & \left(\frac{\ell}{L}\right)^{-(p-1)\frac{\beta}{\nu}}
\qquad \mbox{for $p$ odd}
\end{eqnarray}

For the three-state Potts model such a cancellation does not occur and the
leading behaviour $(\ell /L )^{-p\beta /\nu}$ is affected by a corrective term
as shown in eq. (2.10). However, similar corrections also appear in the
scaling laws of eqs (2.9), which, when taken into account, modify the exact
exponent $\beta
/\nu$ in eq.(2.10). Therefore, for the three-state Potts model, we expect a
more general form
\begin{equation}
F_p(L;\ell )   \sim   \left(\frac{\ell}{L}\right)^{-p\frac{\beta}{\nu}}
\left[A''_p \left(\frac{\ell}{L}\right)^{w} +B''_p\right]
\end{equation}
where $A''_p$, $B''_p$ and $w$ are unknown parameters.\\

{}From this behaviour we deduce the intermittency indices ~: \\

\noindent For the Ising model\\
\begin{eqnarray}
 \lambda_p & = & \frac{1}{8}\; p   \qquad \hspace{1.2cm} \mbox{for $p$ even}
\nonumber\\
	    & = & \frac{1}{8}(p-1) \qquad \mbox{for $p$ odd}
\end{eqnarray}
\noindent For the three-state Potts model\\
\begin{eqnarray*}
 \lambda_p & = & \frac{2}{15}\; p
\end{eqnarray*}

\section{Numerical tests}
We have performed a simulation of the $q=2$ and $q=3$ Potts models at the bulk
critical temperature, using the Swendsen
Wang dynamics, on two-dimensional lattices of size up to $256\times 256$. For
the largest lattice sizes, the data were taken on four independant runs
of $2\times 10^4$ Monte-Carlo lattice updates, with a spacing between two
consecutive measurements of five MC lattice updates. The error analysis is
realised on this total sample of $1.6\times 10^4$ independent
measurements, leading to quite small error bars (of the order of the width of
the data points). In order to
check the reliability of our simulation, we have tested the scaling law for the
order parameter, $<m_q >$ (eq. 2.8), for lattice sizes
ranging from $L=16$ up to $L=256$. The results are displayed in figure~1, with
a linear fit in log scale, giving $\beta /\nu = 0.124(5)$ for the Ising model
(the exact value is 1/8=0.125) and  $\beta /\nu = 0.131(7)$ for the
three-state Potts model (the exact value is 2/15=0.133).\\

We then determine the intermittency indices $\lambda_p$.
We measure the
moments for $p=2$ to $p=5$ and for $\ell =2,4,8,\ldots, L$. Notice
that the values of the moments for $\ell =L$  approach, as
$L\rightarrow\infty$, the ratio of universal
critical amplitudes. For instance, the ratio \mbox{$[F_2(L,\ell
=L)]^2/F_4(L,\ell =L)$}
 has been thoroughly studied~\cite{Binder} and found to have the value
0.85622 in the Ising
model~\cite{Blote}. Our result is 0.856(4) providing a further check of our
data.

In figure 2 we show the moments for the Ising model on a $256\times 256$
periodic lattice.  The behaviour of $F_p$ depending on
the parity of $p$ depicted by eqs (2.11-12) is quite visible on the data.
According to these equations, we have performed a power law fit of the
moments for $4\leq\ell \leq L/2$, shown as the straight lines in figure 2. We
observe that the odd moments are remarkably well fitted by the power law,
whereas the even moments exhibit deviations from this leading behaviour.
Therefore, we
proceed to a double determination of the exponents~: first, from the exact
power law of eqs.(2.11-12) and second, from a corrected parametrisation of the
form of eq.(2.13).
The results of both fitting methods for the largest lattice size $L=256$ are
shown in the last two lines of table I.
The difference between the values corresponding to the same moment
 give an estimate of the systematic error on the exponents. In addition, we
give the result of the power law fit for the $L=128$ lattice, which shows the
stability of this determination.

\begin{center}
\begin{tabular}{||c|c|c|c|c||}
\hline \hline
 & \multicolumn{4}{c|}{\em p}\\ \hline
$ L $ & 2 & 3 & 4 & 5\\
\hline
128 & 0.111 & 0.124 & 0.117 & 0.123\\
\hline
256 & 0.114 & 0.124 & 0.119 & 0.123\\
    & 0.135 & 0.126 & 0.129 & 0.122\\
\hline \hline
\end{tabular}\vspace{0.5cm}	 \\
  \end{center}
\begin{itemize}
\item[\underline{Table I}] The exponents $\lambda_p/p$ ($p$ even) and
$\lambda_p/(p-1)$ ($p$ odd) for the Ising model on
 lattices of size $L=128,256$ and for the moments of order $p=2,3,4,5$.
The exponents given in the bottom line correspond to a fit of
the $L=256$ moments with a corrected scaling law. The expected
exact value is $\beta /\nu =0.125$.\vspace{0.5cm}
\end{itemize}

The agreement of these results with the prediction of eqs.(2.14)
$\beta /\nu =0.125$, is qualitatively good for the even moments and
excellent for the odd ones.

We have repeated the same analysis for
 the three-state Potts model. The moments are shown in figure 3 for the
largest lattice size $L=256$. The fit corresponds to the corrected power law
fit of eq.(2.13)

 $$F_p(L;\ell )=\left(\frac{\ell}{L}\right)^{-px}
\left[ a\left(\frac{\ell}{L}\right)^{w}+b\right]$$
where $x,w,a ,b$ are free parameters.
The results are shown in  table II for the lattice size $L=256$.

 \begin{center}
\begin{tabular}{||c|c|rrrr||}
\hline \hline
L & p  & 2 & 3 & 4 & 5\\
\hline
  & $x$   & 0.141 & 0.124 & 0.131 & 0.125\\
  &	  &$\pm$ 0.015 & $\pm$ 0.009 & $\pm$ 0.009 & $\pm$ 0.011\\
\cline{2-6}
256    & $w$   & 0.359 & 0.270 & 0.307 & 0.235\\
\cline{2-6}
   & $a$ & 0.508 & 0.581 & 0.652 & 0.772\\
\cline{2-6}
   & $b$ & 0.516 & 0.503 & 0.521 & 0.560\\
\hline
128  & $x$   & 0.144 & 0.119 & 0.128 & 0.124\\
  &	  &$\pm$ 0.022 & $\pm$ 0.017 & $\pm$ 0.020 & $\pm$ 0.011\\
\hline \hline
\end{tabular}	  \vspace{0.5cm}   \\
\underline{TABLE II}
\end{center}

\begin{itemize}
\item[\underline{Table II}] The exponents $x=\lambda_p/p$ for the moments of
order $p=2,3,4,5$
for the three-state Potts model on the	$L=256$ lattice. The
expected exact value is $\beta /\nu = 0.133$.
We show the fitted values of the other variables
entering the parametrisation of the  moments
$F_p(z)=z^{-px}(az^{w}+b)$ where $z=\ell /L$. $x$ values from the $L=128$ data
are reported at the bottom line.\vspace{0.5cm}
\end{itemize}

The errors quoted are the  statistical ones corresponding to 95\% confidence
level. They
are quite large due to the freedom allowed by our parametrisation. Actually, we
observe that, if we set
the exponents to the exact value, $x=0.133$, the minimum $\chi^2$ does
not change significantly with respect to its best-fit value.
We give the other variables of our parametrisation  in Table II.
 They are only weakly dependent on the order of the
moment, but the exponents $w$ differ significantly from $x$, indicating, as
expected, a strong contamination of the behaviour of eq.(2.10) by corrections
to the scaling law of eq. (2.9). The exponents obtained from the fit of the
$L=128$ data are reported at the bottom line of table II, showing again the
stability of their determination.

Nevertheless, the agreement with the expected intemittency indices is again
quite good.

\section{Conclusions}
We have defined a block observable for the $q$-state Potts model which exhibits
an intermittent behaviour at the critical point of the model. The moments
of this observable are measurable on finite lattices and have well-defined
scaling behaviour in terms of the block-size. By a numerical simulation, we
confirm the previously observed~\cite{Wosiek,Bambah,Gupta} intermittent
behaviour, and we find
good agreement between the measured exponents and the magnetic exponent $\beta
/\nu$, as predicted by renormalisation group arguments.

Although a critical spin system is presumably not multifractal,
a complete description of its scaling properties requires several exponents.
Besides the thermal and
magnetic critical indices which drive the fluctuations of energy and of
magnetisation, the geometrical aspects of the critical clusters of spins are
characterised by other exponents, not related to the thermodynamics
ones~\cite{Coniglio,VDZande,Knops}. For this reason, the intermittency
phenomenon
in a critical system may assume several aspects depending on the definition of
the block observable.
 This may explain the results of ref~\cite{Gupta},
where the measured intermittency indices coincide with the fractal dimension of
the clusters of spins. However, further theoretical and numerical studies are
needed in order to investigate the geometrical origin of intermittency.\\

\noindent {\Large {\bf Acknowledgements}}\\

\noindent The author thanks R. Peschanski for several valuable discussions and
A. Morel
for his invitation to the Service de Physique Th\'eorique at Saclay (CEA,
France
where this work began. Thanks are also due to E. Pommiers for decisive
contribution in the numerical simulation.

\newpage

\noindent {\Large {\bf Figure Captions}}\\
\begin{itemize}
\item[\underline{Figure 1}] The order parameter for the $q=2$ (Ising) and
$q=3$ Potts
model, measured at the bulk critical temperature as a function of the lattice
size. The lines result from a power fit, giving the exponent $\beta
/\nu=0.124(5)$ for the Ising model (exact value 0.125) and $\beta /\nu
=0.131(7)$ for the three-state Potts model (exact value 0.133).

\item[\underline{Figure 2}] The moments of order $p=2,3,4,5$ for the Ising
model as a
function of the block size~$\ell$. The straight lines result from a power law
fit.
\item[\underline{Figure 3}] The moments of order $p=2,3,4,5$ for the
three-state Potts model as a
function of the block size $\ell$. The curves result from a corrected power law
fit.
\end{itemize}

 \end{document}